\documentclass[twocolumn,aps,prl,superscriptaddress]{revtex4-1}
\usepackage{multirow}
\usepackage{graphicx}
\usepackage{float}
\usepackage{color}
\usepackage{xcolor}
\newcommand{\la}{\langle}
\newcommand{\ra}{\rangle}

\begin{document}


\title{How ``hot precursors'' modify island nucleation: A rate-equations model}

\author{Josue R. Morales-Cifuentes$^{1,2}$}
\email[]{jmorale4@umd.edu, $^\dagger$einstein@umd.edu, $^\ddagger$ap19@rice.edu}
\author{T. L. Einstein$^{1,2},$$^\dagger$}
\author{A. Pimpinelli$^{1,3,\ddagger}$,}
\affiliation{Department of Physics \& $^{2}$CMTC, University of Maryland, College Park, MD 20742-4111 USA \\$^{3}$Rice Quantum Institute \& MSNE Department,  Rice University, Houston, Texas 77005 USA}

\date{\today}

\begin{abstract}
\textcolor{black}{We propose a novel island nucleation and growth model explicitly including transient (ballistic) mobility of the monomers deposited at rate $F$, assumed to be in a hot precursor state before thermalizing. In limiting regimes, corresponding to fast (diffusive) and slow (ballistic) thermalization, the island density $N$ obeys scaling $N \propto F^\alpha$. In between is found a rich, complex behavior, with various distinctive scaling regimes, characterized by effective exponents $\alpha_{\rm eff}$ and activation energies that we compute exactly. Application to $N \left(F, T\right)$ of recent organic-molecule deposition experiments yields an excellent fit.
}
\end{abstract}


\maketitle
The transport of atoms or molecules at solid surfaces plays a crucial role in a huge variety of physical and chemical processes. In heterogenous catalysis reactants usually adsorb and diffuse at the surface before forming products. The growth of regular films implies lateral motion of the adsorbed species. The formation of nanostructures at surfaces via self-assembly or diffusion-limited aggregation (DLA) requires the interplay of the mobility and of the corresponding lateral interactions between constituents. Also, the adsorption process of adspecies may itself involve transient motions.

Curiously,  all of the above but the last have been extensively studied \cite{Stoyanov,Venables,P98,PV,Brune,Mutaf,Oura,EvansThB}; in particular, the effect of ``hot precursors" (transient mobility) on the formation of films and nanostructures has been largely overlooked. The participation of hot precursors to chemical pathways in the formation of molecular hydrogen on interstellar dust grains, or on ice, is considered possible but it is usually neglected \cite{Ofer}.  In descriptions of adsorption, nucleation and growth of thin films and nanostructures, the kinetic energy of the deposited atoms or molecules is commonly assumed to dissipate instantaneously by collisions with surface phonons. Numerical studies \cite{MgO} have shown that this need not be so.  The term ``hot" precursor has been used in surface science for over 3 decades \cite{HarrisKasemo,Tully} to describe transient lateral mobility of adatoms before they chemisorb on the substrate. The concept itself is much older \cite{Kisliuk}.  Widely invoked \cite{Barker,Barth,Keudell}, \textcolor{black}{Ertl's group used hot adatom adsorption over 2 decades ago to account for island nucleation \cite{thatbrunepaper}, followed quickly by Monte Carlo simulations \cite{Albano} and one-dimensional analytics \cite{thetapaper}) but not until recently \cite{Winkler} treated semiquantitatively in the context of island growth.}

In fact, even though the idea that transient, nonthermal motions could be possible after adsorption, when the substrate temperature is low enough, nobody has ever addressed rigorously the modeling of such phenomena. In the present work, we translate that idea into the formalism of rate equations. We describe the deposition of atoms or molecules on a substrate in which the impinging species (``monomers" henceforth) spends some time in a ``precursor'' high-energy, superthermal state in which its motion is ballistic rather than diffusive. The typical lifetime of the precursor state is assumed to be determined by energy exchange between the monomer and the surface phonons.  Consequently, nucleation is diffusion limited at high temperature $T$ and low deposition rate $F$--when energy exchange is favored by abundant phonons, and monomers clusters are far apart---and turns into a novel scaling regime at low $T$ and large $F$.
We show that this novel scenario explains in a natural way recent observations of the puzzling behavior of certain organic molecular adsorbate systems \cite{Winkler}. On more general grounds, we stress that the model proposed in this Letter is sufficiently simple to be amenable to detailed treatment, while exhibiting an unsuspected wealth of different regimes, characterized by a non-monotonic behavior of the activation energies and of the scaling exponents.

We assume that monomers are deposited in a high kinetic energy or \textit{hot} state and propagate ballistically with speed $v$ for a time $\tau_h$, after which they thermalize and diffuse (in random-walk fashion) with diffusion coefficient $D$.  We separate the total monomer density $n$ into the densities of hot $n_h$ and thermalized $n_{th}$ monomers: $n = n_{h} + n_{th}$.  Denoting by $N$ the density of stable (nondecaying) islands, we write the monomer survival times before capture by islands as $\tau_{h \rightarrow N}$ and $\tau_{th \rightarrow N}$, respectively. We then argue for the following rate equations:
\begin{eqnarray}
\label{eqn:original_motion1}
\dot{n}_h    = F - \displaystyle\frac{n_h}{\tau_h} &-& \displaystyle\frac{n_h}{\tau_{h \rightarrow N}}  {\rm;} \quad \quad \dot{n}_{th} = \displaystyle\frac{n_h}{\tau_h} - \displaystyle\frac{n_{th}}{\tau_{th \rightarrow N}} \\
\label{eqn:original_motion2}
\dot{N}      &=& \sigma_i \left(T,v\right) K \left(T,v\right) n^{i+1}
\end{eqnarray}
where $K \left(T,v\right)$ is a kinetic coefficient [see Eq. (\ref{eqn:walton2}) below], $\sigma_s \left(T,v\right)$ is the capture coefficient for a cluster of size $s$, and $i$ is the critical nucleus size, i.e., the size of the largest unstable (decaying) island.
Thus, deposition increases the density of hot monomers while thermalization and island capture decrease it; thermalization is the only source for $n_{th}$.
If we seek only the steady-state scaling behavior, we retain only the dominant terms like monomer capture by stable nuclei ($s \ge i\! +\! 1$); we notably neglect monomer capture by other monomers or by unstable clusters, even including those of size $i$ (cf.~Ref.\cite{Venables}).

We take the associated mean travel distance for both kinds of monomer species to be the mean distance between islands $\bar{\ell}=N^{-1/2}$:
\begin{equation}
\label{eqn:original_motion3}
\tau_{h \rightarrow N}  = \frac{\bar{\ell}}{v} = \frac{1}{v N^{1/2} } {\rm;} \quad \tau_{th \rightarrow N} = \frac{\bar{\ell}\,^2}{D} = \frac{1}{ D N },
\end{equation}

\noindent Our rate equation for the density of clusters of size $s$ is
\begin{equation}
\label{eqn:walton1}
\dot{N}_s = \sigma_{s-1} n N_{s-1} - \sigma_{s} n N_{s} + \frac{1}{\tau_{s+1}} N_{s+1} -\frac{1}{\tau_s} N_s
\end{equation}
where $\tau_s$ is cluster survival time before monomer detachment. In the stationary regime ($\dot{N}_s\! =\! 0$), (\ref{eqn:walton1}) has a unique solution, given the initial condition $N_1 = n$, the so-called Walton relation \cite{Walton}, anticipated by (\ref{eqn:original_motion2}):
\vspace{-0.3cm}
\begin{equation}
N_s = \left( \prod_{k=2}^{s} \sigma_{k-1} \tau_{k} \right) n^s \equiv K_s n^s
\label{eqn:walton2}
\end{equation}
where $K_s$ is the kinetic coefficient for subcritical clusters of size $s$. This necessitates $K_1=1$, which is trivially satisfied under this definition. Comparing the stable island density $N$ in (\ref{eqn:original_motion2}) with its formulation in terms of supercritical clusters, $N \equiv \sum_{s \ge i\! +\! 1} N_s$, we extract the kinetic coefficient in terms of capture and survival times:
\begin{eqnarray}
\dot{N} &=& \sum_{s \ge i\! +\! 1} n \left( \sigma_{s-1} N_{s-1} - \sigma_{s} N_{s} \right) = \sigma_{i} n N_{i} \nonumber \\
\label{eqn:walton3}
&=& \sigma_{i} \left( \prod_{k=2}^{i} \sigma_{k-1} \tau_{k} \right) n^{i+1} \equiv \sigma_{i} K \left(T,v\right) n^{i+1} \; \;
\end{eqnarray}
where (by definition of $i$) $\tau_{s} \rightarrow \infty$ for $s >i$. Note that the kinetic coefficient necessarily depends on the critical cluster size, $K \left(T,v\right)=K_i$. We can obtain explicit expressions for $\sigma_s \left(T,v\right)$ and $K \left(T,v\right)$ for \textit{fast} and \textit{slow} thermalization, viz.\ $\tau_h \ll \tau_{h \rightarrow N}$ and $\tau_h \gg \tau_{h \rightarrow N}$, respectively. Equivalently, if we define
\begin{equation}
z \equiv \tau_h/\tau_{h \rightarrow N} = v \tau_h N^{1/2},
\label{eqn:zdef}
\end{equation}
these limits become $z \! \ll \! 1$ and $z \! \gg \! 1$, respectively.

For fast thermalization, $\tau_h \ll \tau_{h \rightarrow N}$, there are negligibly few hot monomers:  $n \approx n_{th}$. Thermal effects overwhelm ballistic contributions, so $\sigma_s \left(T,v\right) \rightarrow \sigma^{th}_s \left(T\right)$ and $K \left(T,v\right) \rightarrow K^{th} \left(T\right)$, where \textit{th} refers to exclusively thermal contributions. Since the BCF formalism \cite{BCF} applies,
\begin{eqnarray}
\label{eqn:fast_coefficients}
\sigma^{th}_s \left(T\right)= D = D_0 e^{-\beta E_D} {\rm;} \quad \quad K^{th}(T) = \kappa_0 e^{\beta E_i}
\end{eqnarray}
where $E_D$ is the diffusion energy, $E_i$ is the cohesion energy of a cluster of size $i$, $\beta \equiv (k_BT)^{-1}$ is the inverse thermal energy, and $D_0$ and $\kappa_0$ are constants. It is simpler to work in terms of the coverage $\theta \equiv F t$, so $\dot{N} = F dN/d\theta$. In the stationary regime, $\dot{n}_h = \dot{n}_{th} = 0$, (\ref{eqn:original_motion1}) simplifies:
\begin{eqnarray}
\label{eqn:fast_motion1}
n_h    = F \tau_h {\rm;} \quad  n_{th} = \displaystyle\frac{F}{D N}{\rm;} \quad
\frac{d N}{d \theta} = \kappa_0 \left( \frac{F}{D} \right)^{i} \frac{e^{\beta E_i}}{N^{i+1}}. \;
\end{eqnarray}
In the $z \ll 1$ regime we find the familiar case of DLA: integrating $dN/d\theta$ yields the DLA hallmarks \cite{Venables,Stoyanov,VPW,P98}:
\begin{equation}
N \propto  \left(\frac{F}{D_0} \right)^{\alpha} \exp \left[\beta  \frac{i E_D + E_i}{i+2} \right]; \;\;\; \alpha = \displaystyle{i}/({i+2})
\label{eqn:fast_motion2}
\end{equation}

For slow thermalization,  $z \gg 1$, we find a novel hot monomer aggregation (HMA) regime: Since our goal is to understand the scaling behavior of $N_s$ rather than its distribution, we neglect the much-studied (and reviewed \cite{Venables,EvansThB,Einax}) dependence of $\sigma_s$ on $s$. We focus on the effect on scaling of the domination by hot monomers moving ballistically at some hyperthermal speed $v$, taking for dimensional reasons
$\sigma_s^B = \ell v$, where the coefficient 
$\ell$ is a characteristic microscopic length and \textit{B} refers to this ballistic regime. A hot monomer colliding with a small cluster is likely to transfer energy to the latter and cause (thermally improbable) detachment of a previously attached monomer. Adopting the simplest assumption that this athermal detachment rate is proportional to the monomer speed, we get the cluster lifetime $\tau_s^B=\ell'/v$, where $\ell'$ is another microscopic length.
\begin{equation}
K_s^B = \left( \prod_{k=2}^{s} \sigma_{k-1}^B \tau_{k}^B \right) = \left( \ell \ell ' \right)^{s-1}
\label{eqn:slowkinetic}
\end{equation}
The noteworthy independence of $K_s^B$ on  $v$ is robust, requiring only that $\sigma_s^B$ and $(\tau_s^B  )^{-1}$ have the same speed dependence; if not linear, some of the dimensionless quantities defined below are trivially rescaled.  Even if $\sigma_s^B$ and $(\tau_s^B)^{-1}$ had different speed dependences, both would still be independent of $T$ and $F$, so that the effective exponents and energies in Table \ref{tab:exponents12}  would not change \cite{supplement}.

Since $n \approx n_{h}$, in the steady state:
\begin{eqnarray}
\label{eqn:slow_motion1}
n_h \! = \! \frac{F}{v N^{1/2}} {\rm ;} \quad n_{th}  \! = \!  \frac{F\tau_h^{-1}}{D  v N^{3/2}} {\rm ;} \quad
\frac{d N}{d \theta}  \! = \!  \frac{\ell K_i^B	}{N^{\frac{i+1}{2}}} \left( \frac{F}{v} \right)^{i},
\end{eqnarray}
which remarkably leads to the scaling form $N \propto F^\alpha$ with
\begin{equation}
\label{HMA}
\alpha = 2i/(i+3)
\end{equation}
the same scaling exponent as attachment-limited aggregation (ALA) \cite{Kandel,VB,PTW}. Contrary to ALA, which depends on thermally activated processes, the novel HMA nucleation regime is essentially athermal, evident from the temperature independence of the coefficients of Eq.~(\ref{eqn:slow_motion1}).

To solve the model, redefine the capture coefficients and monomer survival times as $\tilde{\sigma}_s \left(T,v\right) $ and $\tilde{\tau}_s \left(T,v\right)$:
\begin{eqnarray}
\label{eqn:thermal1}
n \tilde{\sigma}_s \left(T,v\right) &=& n_{th} \sigma^{th}_s \left(T\right) + n_{h} \sigma^{B}_s \left(v\right), \\
\label{eqn:thermal2}
\frac{1}{\tilde{\tau}_s \left(T,v\right)} &=& \frac{1}{\tau^{th}_s \left(T\right)} + \frac{1}{\tau^{B}_s \left(v\right)},
\end{eqnarray}
\noindent where thermal and ballistic contributions contribute linearly for the capture coefficients and additively for the survival rates. This recasts (\ref{eqn:walton1}) as:
\begin{equation}
\dot{N}_s = \tilde{\sigma}_{s-1} n N_{s-1} - \tilde{\sigma}_{s} n N_{s} + \frac{1}{\tilde{\tau}_{s+1}} N_{s+1} -\frac{1}{\tilde{\tau}_s} N_s
\label{eqn:waltonGeneral}
\end{equation}
At stationarity we recoup (\ref{eqn:original_motion1}) and, with $\sigma_s \rightarrow \tilde{\sigma}_s$ and $\tau_s \rightarrow \tilde{\tau}_s$ we recover (\ref{eqn:walton3}). Unsurprisingly, from our assumptions for the slow thermalization, the $s$ dependence of $\tilde{\sigma}_s$ and $\tilde{\tau}_s$ can be neglected. Two undetermined survival times remain: $\tau_h$ for molecules before thermalization and $\tau^{th}$ for thermalized monomers before adsorption into an island. In accordance with the BCF model \cite{BCF}, both are assumed of the form
\begin{equation}
\label{eqn:scales1}
\tau^{th}_s \equiv \tau_0 e^{\beta E_b} {\rm ,} \quad \quad \tau_h \equiv \tau_0 e^{\beta E_{ph}},
\end{equation}
\noindent where $\tau_0$ is a characteristic inverse phonon frequency, $E_b$ is the barrier to detachment of a monomer from a cluster ($E_b > E_D$), and $E_{ph}$ is a typical phonon energy.

Using the algebraic forms for $n_h$ and $n_{th}$ in (\ref{eqn:original_motion1}) and the survival time considerations in (\ref{eqn:original_motion3}), we solve (\ref{eqn:thermal1}) for the capture coefficient in steady state:
\begin{eqnarray}
\label{eqn:exact1}
n_h &=& \displaystyle\frac{ F \tau_h }{ 1 + v \tau_h N^{1/2} }  =  N D \tau_{h} n_{th} \\
\label{eqn:exact2}
\tilde{\sigma} &=& \displaystyle\frac{n_h}{n \tau_{h} N} \left( 1 + \ell v \tau_{h} N \right) = D \left( \frac{1+ \ell v \tau_{h} N}{1 + N D \tau_{h}} \right),
\end{eqnarray}
\noindent Substituting into the recast form for (\ref{eqn:walton3}) we obtain
\begin{equation}
\frac{dN}{d \theta} = \frac{   \left( \tau F \right)^{i} }{D  \tau }  \frac{  \displaystyle\left(   1 + N D \tau_{h} \right) \displaystyle\left( 1 + \ell v \tau_{h} N \right)^{i} }{   N^{i+1} \displaystyle\left( 1 + v \tau_h N^{1/2} \right)^{i+1}}
\label{eqn:exact3}
\end{equation}
\noindent using the approximation $\ell \!  =\! \ell^\prime \!$.  In the limits $z \ll 1$ and $z \gg 1$, we recover, to leading order, the key characteristics of $dN/d\theta$  shown in (\ref{eqn:fast_motion2}) and (\ref{HMA}), respectively: the dependence on $F$, $N$ and the DLA and HMA/ALA scaling exponents.
Note that (\ref{eqn:exact3}) is a rational function of $N^{1/2}$ (evident after a change of variables from $N$), so that it is analytically solvable using partial-fraction decomposition for arbitrary values of $i$; however, the result is both unwieldy and unenlightening.  Instead, we adopt a scaling approach to study $N$ vs.\ six dimensionless parameters that can be formed from the physical quantities $\theta$, $F$, $D$, $\tau_h$, $v$, and $\tilde\tau$:
\begin{eqnarray}
\label{eqn:dimless}
\hat N \equiv N(v\tau_h)^2 = z^2; \;\;\; \hat F &\equiv& F \ell v \tau_h^2; \;\;\; \hat{\theta} \equiv \frac{\theta v}{D \ell}  (v\tau_h)^4  \quad \quad \\
\mathcal{R}_{C} \equiv \frac{\ell}{v \tau_0} e^{- \beta E_b} 	;\;\; \mathcal{R}_{n} &\equiv& N D \tau_{h} ; \;\;  \mathcal{R}_B \equiv \ell v \tau_{h} N  \: \nonumber
\end{eqnarray}
We can then solve (\ref{eqn:exact3}) implicitly:
\begin{eqnarray}
\frac{\hat{\theta} \, \hat{F}^i}{ \left( 1 + \mathcal{R}_C \right)^{i-1}} =\! \int_{0}^{\hat{N}}\!\!\! f(\epsilon) \, d \epsilon \! {\rm ,} \:\:\:\: f(\epsilon)\! \equiv \! \frac{ \epsilon^{i\! +\! 1} \left( 1 \! +\! \epsilon^{1/2} \right)^{i\! +\! 1} }{ \left( 1 \! +\!  R_n \epsilon \right) \! \left( 1\! +\!  R_B \epsilon \right)^{i}} \: \nonumber \\
\label{eqn:sol1}
\end{eqnarray}
where we introduce $R_n \equiv \mathcal{R}_n / \hat{N}$ and $R_B \equiv \mathcal{R}_B / \hat{N}$ to easily identify prefactors of $\hat{N}$ within numerical computations\textcolor{black}{, and $\epsilon$ is an integration variable.}
\begin{figure}[t]
\includegraphics[scale=0.38]{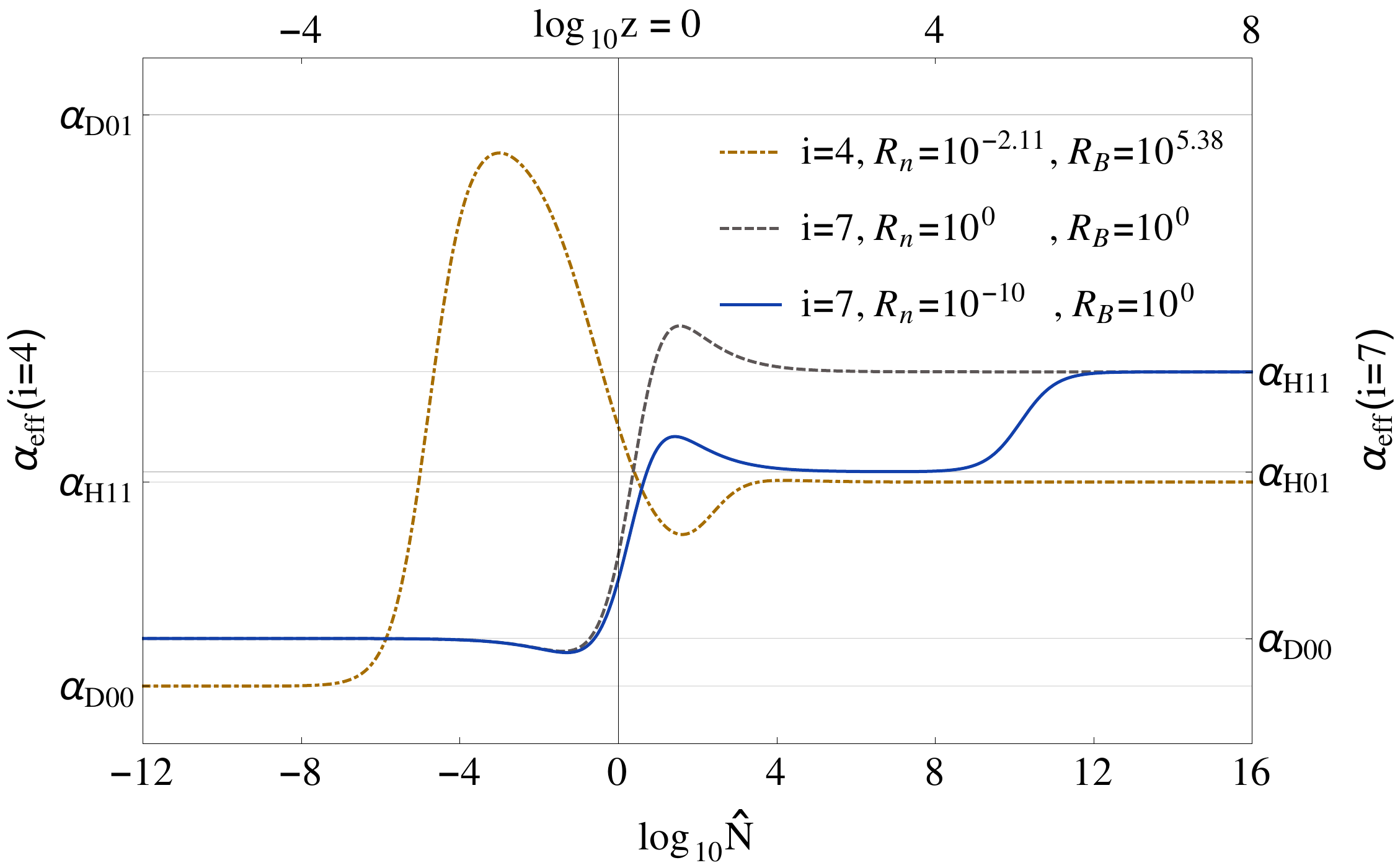}
\caption{\label{fig:exponents1} The effective exponent $\alpha_{\rm eff}$ vs.\ $\hat{N}(F)$ for $i=4$ (dash-dotted line) and 7 (dashed and continuous lines). Note the different scales for $\alpha_{\rm eff}$ depending on $i$. The crossover region between the limiting DLA (D00) and HMA (H11) scaling is explored by varying  $R_{n}$ and $R_{B}$. All eight possible curves are plotted, at fixed $i$, in \cite{supplement}.
}
\end{figure}

For both the fast and slow thermalization limits, $N \propto F^\alpha$ has a well-defined power-law exponent $\alpha$. We define an \textit{effective exponent} $\alpha_{\rm eff}(F) \equiv d \ln N/d \ln F$ between the scaling limits. From (\ref{eqn:sol1}) follows the explicit form of $\alpha_{\rm eff}$:
\begin{equation}
\label{eqn:alpha}
\alpha_{\rm eff}\left( F \right)  = \frac{i \int_{0}^{\hat{N}} \! f(\epsilon)\, d \epsilon}{\hat{N}f(\hat{N})} = i \left<1\right> \left|_{\hat{N}} \right.
\end{equation}
where $\alpha_{\rm eff}$ depends implicitly on $F$ via $\hat{N}$. We also employed the notational shorthand, $\left< \cdots \right>$:
\begin{equation}
\label{eqn:transform}
\left< g\right> \left|_{\hat{N}} \right. \equiv  \left[ \hat{N} f(\hat{N}) \right]^{-1} \int_{0}^{\hat{N}} \!\!\! f(\epsilon)  g(\epsilon) \, d\epsilon .
\end{equation}
\textcolor{black}{For present purposes we can unambiguously omit the $\hat{N}$ indexation} from the notation. By varying $\mathcal{R}_n$ and $\mathcal{R}_B$, we can explore the various regimes, as exemplified in Fig.~\ref{fig:exponents1}. We see that $\alpha_{\rm eff}$ always converges to its limiting DLA (HMA) values for small (large) $\hat N$---or, equivalently, for small(large) $F$ or $z$.  However, the crossover behavior exhibits nontrivial features: $\alpha_{\rm eff}$ may lock into plateaus of rational values, over which the island density exhibits well-defined power-law behavior (cf.\ continuous line in Fig. 1). With Taylor expansions of (\ref{eqn:alpha}) in $\hat{N}$ or $\hat{N}^{-1}$, such values are found analytically and given in Table \ref{tab:exponents12}. Especially interesting, and relevant to experiments as discussed below, is the behavior for $i=4$ in Fig. 1. Here $\alpha_{\rm eff}$ has a large maximum nearly equal to $\alpha_{\rm D01}=2$, between $\alpha_{\rm D00}=2/3$ and $\alpha_{\rm H11}=8/7$. Measuring the island density between, for instance, $z \approx10^{-4}$ and $z\approx10^{-2}$ could be interpreted as a transition between  $\alpha_{\rm D00}=7/9$ and $\alpha_{\rm H11}=7/6$ with $i=7$.

Since we also expect $N \propto \exp(\beta E_A)$ in simple limiting cases where there is a well-defined activation energy $E_A$, we can seek an \textit{effective activation energy} $E_A^{\rm eff} = d \ln N/d\beta$ for intermediate situations. After long algebraic manipulations, we find a compact form for the activation energy:
\begin{eqnarray}
E_A^{\rm eff} &=& E_b (i\! -\! 1) \frac{\mathcal{R}_C}{1 + \mathcal{R}_C} \left< 1 \right> - E_{ph} \left(i+1\right) \left< \frac{z}{1+z} \right> \nonumber \\
&+& i E_{ph} \left< \frac{\mathcal{R}_B}{1+\mathcal{R}_B} \right>+ \left< \frac{E_D}{1+\mathcal{R}_n} + \frac{E_{ph} \mathcal{R}_n}{1+\mathcal{R}_n} \right> ,
\label{eqn:act2}
\end{eqnarray}
\begin{table}[t]
\centering
\begin{tabular}{ c || c | c | c || c || c }
\hline
\hline
Label & $\mathcal{R}_n$  & $\mathcal{R}_B$ & $\mathcal{R}_C$ & $\alpha_{\rm eff} = i\la 1 \ra$ & $E_A^{\rm eff}/\la 1 \ra = E_A^{\rm eff} i/\alpha_{\rm eff}$ \\
\hline
\hline
D$_{000}$ & $\ll \! 1$ & $\ll \! 1$ & $\ll \! 1$ & $i/(i \! + \! 2)$    & $E_D                                             $ \\
D$_{001}$ & $\ll \! 1$ & $\ll \! 1$ & $\gg \! 1$ & \texttt{"}           & $E_D \! + \! (i \! - \! 1)E_b                    $ \\
D$_{010}$ & $\ll \! 1$ & $\gg \! 1$ & $\ll \! 1$ & $i/2$                & $E_D \! + \! i E_{ph}                            $ \\
D$_{011}$ & $\ll \! 1$ & $\gg \! 1$ & $\gg \! 1$ & \texttt{"}           & $E_D \! + \! i E_{ph} \! + \! (i \! - \!1)E_b    $ \\
D$_{100}$ & $\gg \! 1$ & $\ll \! 1$ & $\ll \! 1$ & $i/(i \! + \! 1)$    & $E_{ph}                                          $ \\
D$_{101}$ & $\gg \! 1$ & $\ll \! 1$ & $\gg \! 1$ & \texttt{"}           & $E_{ph} \! + \! (i \! - \! 1)E_b                 $ \\
D$_{110}$ & $\gg \! 1$ & $\gg \! 1$ & $\ll \! 1$ & $i$                  & $(1 \! + \! i ) E_{ph}                           $ \\
D$_{111}$ & $\gg \! 1$ & $\gg \! 1$ & $\gg \! 1$ & \texttt{"}           & $(1 \! + \! i ) E_{ph} \! + \! (i \! - \! 1)E_b  $ \\
\hline
H$_{000}$ & $\ll \! 1$ & $\ll \! 1$ & $\ll \! 1$ & $2i/\left(3i \! + \! 5\right) $ & $E_D \! - \! (i \! + \! 1)E_{ph}                         $ \\
H$_{001}$ & $\ll \! 1$ & $\ll \! 1$ & $\gg \! 1$ & \texttt{"}                      & $E_D \! - \! (i \! + \! 1)E_{ph} \! + \! (i \! - \! 1)E_b$ \\
H$_{010}$ & $\ll \! 1$ & $\gg \! 1$ & $\ll \! 1$ & $2i/\left(i \! + \! 5\right)  $ & $E_D \! - \! E_{ph}                                      $ \\
H$_{011}$ & $\ll \! 1$ & $\gg \! 1$ & $\gg \! 1$ & \texttt{"}                      & $E_D \! - \! E_{ph} \! + \! (i \! - \! 1)E_b             $ \\
H$_{100}$ & $\gg \! 1$ & $\ll \! 1$ & $\ll \! 1$ & $2i/\left(3i \! + \! 3\right) $ & $-i E_{ph}                                               $ \\
H$_{101}$ & $\gg \! 1$ & $\ll \! 1$ & $\gg \! 1$ & \texttt{"}                      & $-i E_{ph} \! + \! (i \! - \! 1) E_b                     $ \\
H$_{110}$ & $\gg \! 1$ & $\gg \! 1$ & $\ll \! 1$ & $2i/\left(i \! + \! 3\right) $  & $0                                                       $ \\
H$_{111}$ & $\gg \! 1$ & $\gg \! 1$ & $\gg \! 1$ & \texttt{"}                      & $(i \! - \! 1)E_b                                        $ \\
\hline
\hline
\end{tabular}
\caption{\textcolor{black}{The 16 regimes for extremal values of $z$, $\mathcal{R}_n$, $\mathcal{R}_B$ and $\mathcal{R}_C$, along with the associated rescaled effective exponents (cf.\ Fig.~\ref{fig:exponents1}) and effective activation energies. D (H) indicates DLA (HMA): $z \ll (\gg) 1$. The subscripts give the limiting value of the $\mathcal{R}$ 's, with 1 denoting $\mathcal{R}\! =\! \infty$, i.e. index = $\exp(-1/\mathcal{R})$.  For $\alpha_{\rm eff}$, $\mathcal{R}_C$ is inconsequential, so that only the first 2 subscripts are needed,} yielding just 8 regimes.   Note that the reduced values of $E_A^{\rm eff}$ in the last column must be multiplied by the corresponding $\alpha_{\rm eff}/i$ to get the actual $E_A^{\rm eff}$.}
\label{tab:exponents12}
\end{table}
\begin{figure}[t]
\includegraphics[scale=0.38]{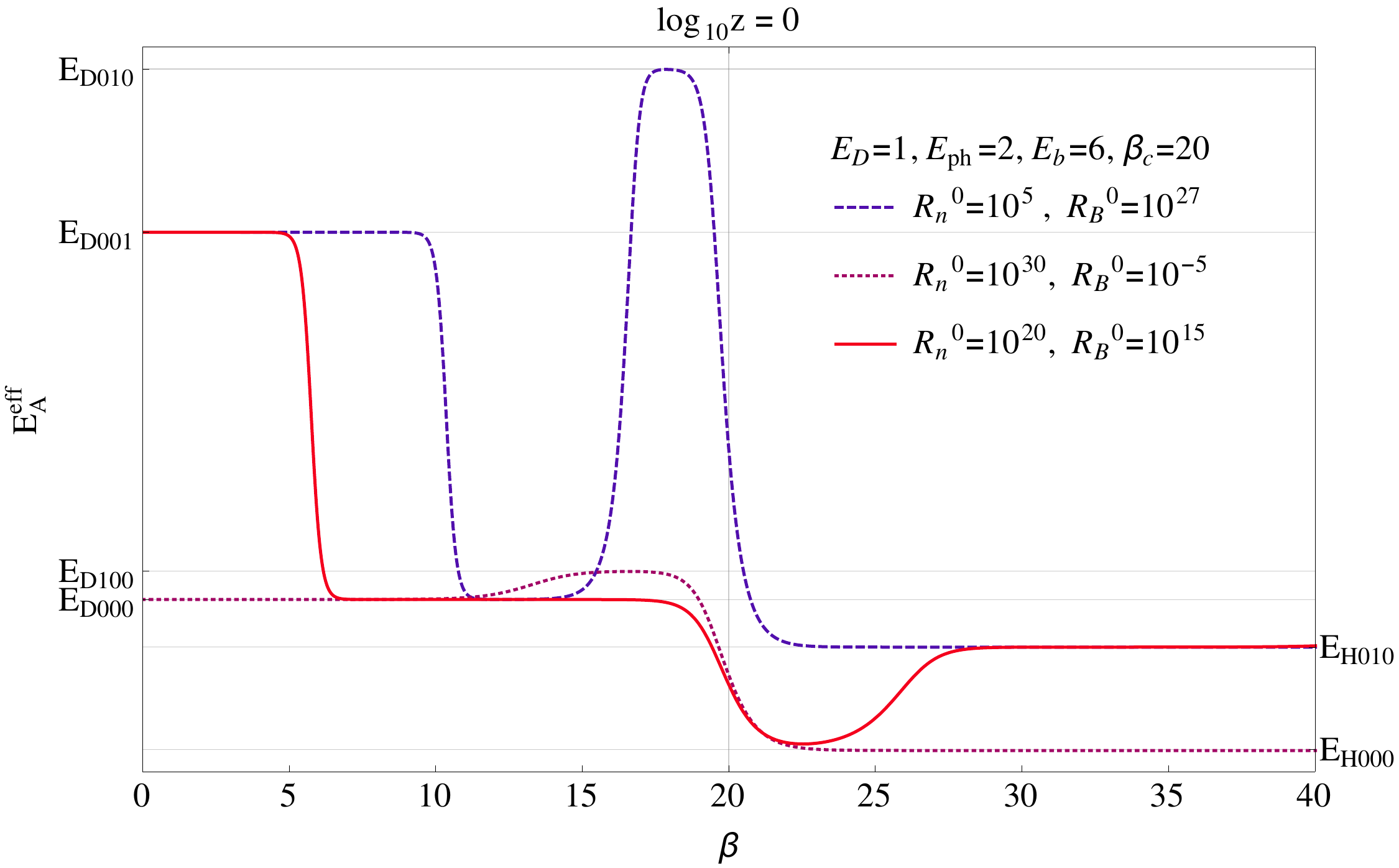}
\caption{
\textcolor{black}{$E_A^{\rm eff}$~vs.~$\beta$~for~$i\! =\! 4$~and~some~values~of~$R_n^0$~and~$R_B^0$,~$R_x^0$ meaning $R_x(\beta\! =\! 0)$.  The fast (slow) thermalization regimes are left (right) of the crossover value $z\! =\! 1$: to satisfy $z(\beta_c\! =\! 20)\! =\! 1$ each curve needs a different value of $\hat{\theta}^0=\hat{F}^0$. The nonmonotonic crossover and associated well-defined plateaus are sensitive to $E_D$, $E_{ph}$, $E_b$, $\beta_c$, $R_n^0$, and $\! R_B^0$, leading to a rich variety of behaviors, presented more fully in \cite{supplement}}.}
\label{fig:arrfast}
\end{figure}
\noindent \hspace{-1mm}\textcolor{black}{which uses the bracket notation from (\ref{eqn:transform}), while $z$, $\mathcal{R}_C$, $\mathcal{R}_B$, and $\mathcal{R}_n$ remain functions of $\hat{N}$ \cite{constrain}.} (See \cite{supplement} for a full derivation.)  There are $2^4=16$ regimes realizable by varying $z$, $\mathcal{R}_n$, $\mathcal{R}_B$ and $\mathcal{R}_C$: their effective energies $E_A^{\rm eff}$ are in Table \ref{tab:exponents12}, with selected plots in Fig.\ \ref{fig:arrfast}.

\begin{figure}[b]
\includegraphics[scale=0.28]{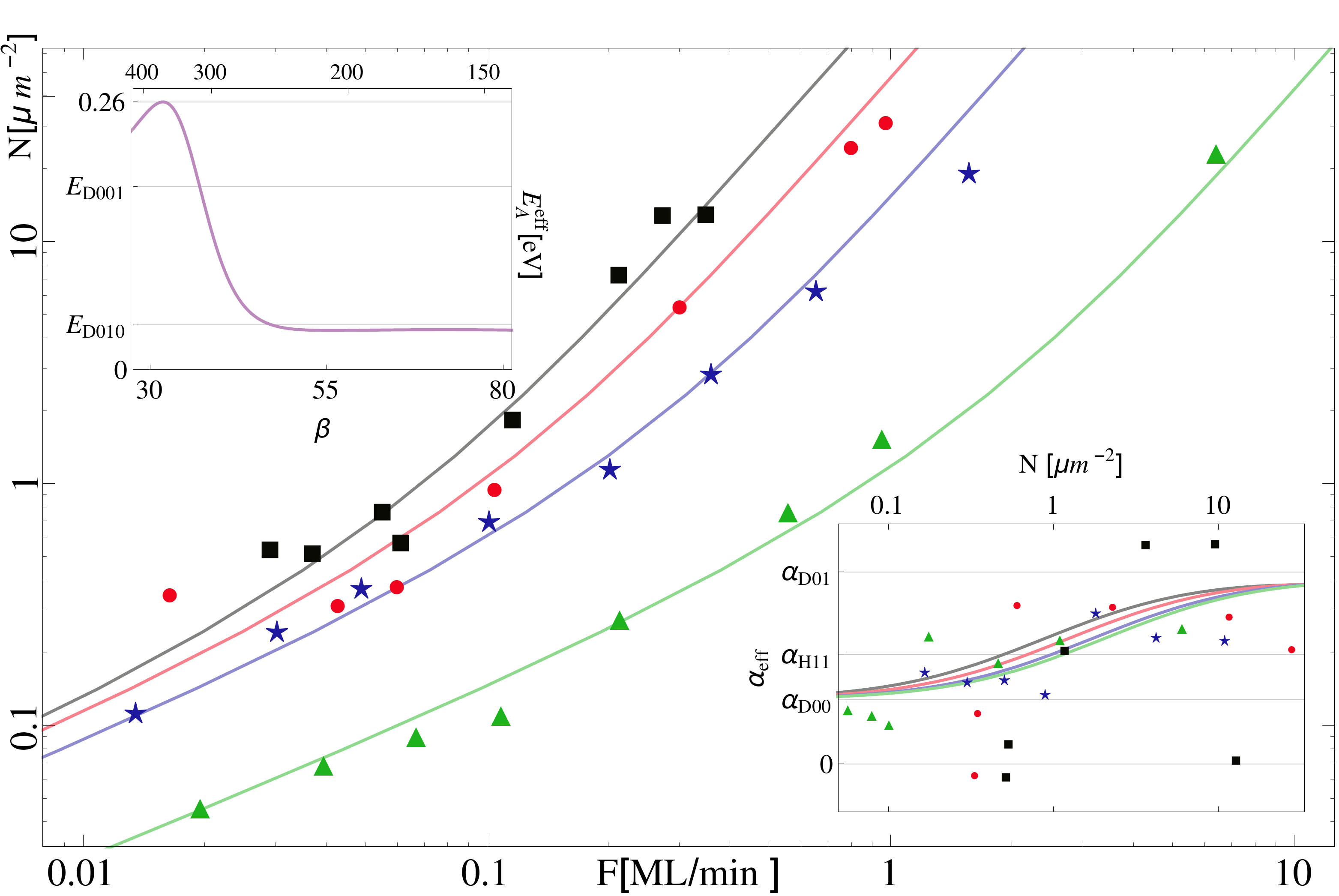}
\caption{\label{fig:experiment} Island density vs.\ deposition rate in the aggregation regime for 6P on sputter-modified mica (001) \cite{tumbekexp} at $T \! = \! 150$K (\textit{black}, square dots), $200$K (\textit{red}, round dots), $300$K (\textit{blue}, star dots), and $400$K (\textit{green}, triangular dots), with best-fit parameters: $i \! = \! 4$, $\log_{10} \mathcal{R}_n^0 \! = \! -2.11$, $\log_{10} \mathcal{R}_B^0 \! = \! 5.38$, $\log_{10} v \tau_0 \left[\mu m\right] \! = \! -2.66$, and, in eV, $E_D \! = \! 0.0174$, $E_{ph} \! = \! 0.0174$, and $E_b \! = \! 0.349$. The final effective coverage $\theta_{\rm eff}$ \cite{fitdetails} is given by $\log_{10} \left( \theta_{\rm eff} \tau_0^i \left[\mu m^{-2}s^{i}\right] \right) \! = \! -12.4$ . \textit{Inset, bottom right}, $\alpha_{\rm eff}$ (curves) and values of $d \ln N/d \ln F$ from the experimental data (dots); \textit{Inset, top left,} $E^{\rm eff}_{A}$ (curve) vs.\ 1/$T$[K] for $F = 0.55$ML/min.
}
\end{figure}

In the regime $z \gg 1 $ the values of $E_A^{\rm eff}$ can be negative \cite{neg}.  The observation of $E_A^{\rm eff}\! <\! 0$ in similar surface experiments \cite{gottfried2003, matetsky2013, holloway1974} was attributed to a Langmuir-Hinshelwood mechanism \cite{lazaga, minnegative}.  Here the key phenomenon is the onset of long-distance ($\gg \ell^\prime$) ballistic motion with decreasing $T$ that competes with diffusive aggregation to reduce $N$ over a range of $T$  \cite{supplement}.
Further insight into $E_A^{\rm eff}$ is gained by interpreting the limiting values of the $\mathcal{R}$'s \cite{supplement}, e.g., for $\mathcal{R}_n= n_h/n_{th} \gg 1$ (hot-precursor domination), the last term of (\ref{eqn:act2}) tends to $E_{ph}$, while for $\mathcal{R}_n \ll 1$ it goes to $E_D$, consistent with domination by diffusing thermal adatoms.

To test our model, we fit the experimental island density of hexaphenyl (6P) deposited on sputter-modified mica at $T=150$, $200$, $300$, and $400$K \cite{tumbekexp}.
The fit, shown in
Fig.~\ref{fig:experiment}, yields $i \! = \! 4 \pm 1$ and strongly suggests that the system explores the crossover region between $D_{000}$ and $D_{010}$, i.e. $-3.3 \leq \log_{10} z \leq -2.0$ \cite{fitdetails}. The caption of Fig. 3 lists the parameters. The bottom right inset, showing the numerical derivative of the experimental data along with $\alpha_{\rm eff}$ via (23), confirms that $\alpha_{\rm eff}$ varies from $\alpha_{\rm D00}$ to $\alpha_{\rm D01}$ and shows the insensitivity of $\alpha_{\rm eff}$ to $T$. Previous estimates of $i$ \cite{tumbekexp}, which did not envision nonmonotonic crossover, assumed DLA scaling at low $F$ and  ALA scaling at high $F$ and so found a varying critical-nucleus size, viz. $i=5 \pm 2$ and $i \! = \! 7 \! \pm \! 2$, respectively \cite{tumbekexp}; mistaking the novel $D_{000} \rightarrow D_{010}$  transition for $D_{000} \rightarrow H_{110}$ (where $\alpha_{\rm H110}=\alpha_{\rm ALA}$). The top left inset shows that the computed $E_A^{\rm eff}$ varies in this $T$ range between ~0.26 eV (high-$T$) and ~0.04 eV (low-$T$), corresponding remarkably well to the experimental values of 0.3 eV and 0.04 eV, respectively \cite{Winkler}.  This highlights the success of our hot-precursor model to account for crossover behavior in nucleation. The model can be extended to consider chemical reactions in the context of deposition of different chemical species. A similar model was proposed by one of the authors, several years ago, to describe precursors in chemical vapor deposition \cite{TP}.


Our ground-breaking work shows systematically that nonthermal (high-energy) adsorption states have dramatic effects on monomer aggregation and island nucleation: while known scaling relations in limiting situations are recovered, a novel scaling regime is discovered, as well as intermediate regimes in which different scaling behaviors occur for well-defined ranges of the controlling parameters. Far from being a tweak, ``hot'' monomers profoundly modify island nucleation, begging further theoretical and experimental investigation. Our work constitutes the first, crucial step in that direction. While a detailed confrontation of a specific system, scrutinizing the many input parameters of our model, will require more extensive experimental data,
the preliminary test of our model is encouragingly consistent with the best available data \cite{tumbekexp} and captures the main physical aspects of the nucleation of molecular thin films, presaging enormous technological potential \cite{OTFT}.


\vspace{-5mm}
\section*{Acknowledgements}
\vspace{-3mm}
Work at UMD supported by NSF Grant No.~CHE 13-05892.  A.P. thanks A. Winkler for inspiring discussions.


\begin{thebibliography}{99}
\bibitem{constrain} \textcolor{black}{To ensure that $z(\beta \rightarrow 0) \rightarrow 0$, an adequately small value for $\hat{\theta} \hat{F}^i$ is required.  In the limit $\beta \rightarrow \infty$, (\ref{eqn:sol1}) implies $\hat{N} \rightarrow \infty$ if $E_D > E_{ph}$, which would constrain the system to $E_{ph} \geq E_D$ if the model were still valid.}
\bibitem{neg} Especially for $E_{ph}\! >\! E_B$, but not when $\mathcal{R}_n\!  \gg\! 1, \mathcal{R}_B\!  \gg\! 1$.
\bibitem{fitdetails} We assume all physical parameters remain constant for every $T$. As the final coverage varies over the range 0.1-0.3 ML for every measurement, the analysis refers to an effective final coverage $\theta_{\rm eff}$ for the system.  As $\alpha_{\rm eff}$ transits $\alpha_{\rm D00} \rightarrow \alpha_{\rm D01}$, $-9.3 \leq \log_{10} \mathcal{R}_n \leq -6.6$, $-6.4 \leq \log_{10} \mathcal{R}_B \leq 1.0$, and $-1.5 \leq \log_{10} \mathcal{R}_C \leq 1.1$.
\end{thebibliography}
\end{document}